\let\nopictures=N
\documentstyle[12pt]{article}
\newcommand{\be}{\begin{equation}}
\newcommand{\ee}{\end{equation}}
\newcommand{\bea}{\begin{eqnarray}}
\newcommand{\eea}{\end{eqnarray}}
\newcommand{\nen}{\nonumber \\ \relax}

\newcommand{\forcepar}{{\hskip 10pt\vskip -15pt}}
\newfont{\headfont}{cmbx10 scaled 1440}
\newfont{\namefont}{cmr10}
\newfont{\initialfont}{cmr10 scaled 1200}
\newfont{\addfont}{cmti10}
\newfont{\boldmathfont}{cmbx10}
\newfont{\figfont}{cmr7 scaled 1200}

\newcommand{\seq}{\ =\ }
\newcommand{\pls}{\ +\ }
\newcommand{\mi}{\ -\ }
\newcommand{\seqv}{\ \equiv\ }
\newcommand{\half}{\frac{1}{2}}
\newcommand{\inv}[1]{\frac{1}{#1}}

\newcommand{\pathint}{{\boldmathfont\int}}

\newcommand{\ca}{{\cal A}}

\newcommand{\cc}{{\cal C}}

\newcommand{\cf}{{\cal F}}
\newcommand{\cg}{{\cal G}}

\newcommand{\co}{{\cal O}}
\newcommand{\cs}{{\cal S}}

\newcommand{\cv}{{\cal V}}

\newcommand{\IR}{{I \kern -0.4em R}}
\newcommand{\IC}{{I \kern -0.65em C}}

\newcommand{\dimm}{D}
\newcommand{\rank}{(\dimm-1)}
\newcommand{\dual}[1]{{^*\! #1}}
\newcommand{\mywedge}{\leavevmode{\raise -.4ex\hbox{$\,\wedge\,$}}}
\newcommand{\ldim}[1]{\ell{\rm dim}[#1]}
\newcommand{\lap}{\Delta}

\newcommand{\tq}{-\inv{\tau}}

\newcommand\mybox[2]{\vcenter{\hrule width#1in\hbox{\vrule height#2in
   \hskip#1in\vrule height#2in}\hrule width#1in}}
\newcommand\dal{{\mybox{.1}{.1}}}

\newcommand{\ev}[1]{\langle #1 \rangle}

\newcommand{\khat}{\hat K}
\newcommand{\ohat}[1]{{\hat \co}_{#1}}
\newcommand{\seqarrow}{\hbox to 27pt{\rightarrowfill}}
\newcommand{\sar}{\ \seqarrow\ }
\newcommand{\oparr}[1]{\ \buildrel {#1}\over \seqarrow\ }

\begin{document}
\begin{titlepage}
\renewcommand{\thefootnote}{\fnsymbol{footnote}}
\begin{center}
{\headfont The Cosmological Constant and\break Volume--Preserving
Diffeomorphism Invariants}\footnote{This work is supported in part by funds
provided by the
U. S. Department of Energy (D.O.E.) under contract
\#DE-AC02-76ER03069.}

\end{center}
\vskip 0.3truein
\begin{center}
{{\initialfont R}{\namefont OGER}
		    {\initialfont B}{\namefont ROOKS}}
\end{center}
\begin{center}
{\addfont{Center for Theoretical Physics,}}\\
{\addfont{Laboratory for Nuclear Science}}\\
{\addfont{and Department of Physics,}}\\
{\addfont{Massachusetts Institute of Technology}}\\
{\addfont{Cambridge, Massachusetts 02139 U.S.A.}}
\end{center}
\vskip 0.5truein
\begin{abstract}

Observables in the quantum field theories of  $(D-1)$-form fields, $\ca$, on
$D$-dimensional, compact and orientable manifolds, $M$, are computed.
Computations of the vacuum value of $T_{ab}$ find it to  be the metric times
a function of the volume of spacetime, $\Omega(M)$.  Part of this function of
$\Omega$ is a finite zero-mode contribution.  The correlation functions of
another
set of operators give intersection numbers on $M$.  Furthermore, a similar
computation for products of  Wilson area operators results in  a function of
the volumes of the intersections of the submanifolds the
operators
are defined on.  In addition, scalar field couplings are introduced and
potentials are  induced after integrating out the $\ca$ field.
Lastly, the thermodynamics of the pure theories is found to be analogous to the
zero-point motion of a scalar particle.  The coupling  of a Gaussian scalar
field to the $\ca$ field is found to manifest itself on the free energy at high
temperatures and/or small volumes.
\vskip 0.5truein
\leftline{CTP \# 2247  \hfill September 1993}
\smallskip
\leftline{hep-th/9310007}
\end{abstract}

\end{titlepage}

\section{Introduction}\label{INTRO}
\forcepar
Theories of higher forms were largely spawned by supergravity investigations.
This is especially true of 3-form theories.  Supergravity in eleven dimensions
contains such a form \cite{REFCJS}.    As a result of the
presence of the 3-form in
its eleven-dimensional parent theory, $D=4$, $N=8$  supergravity  also
has a 3-form in its
field content.   In addition, a formulation of D=4, N=1
supergravity
exists in which a 3-form appears \cite{REFFDSG}.
One interesting aspect of the four dimensional 3-form theory is that it gives
a
contribution to the cosmological constant which arises purely as an
integration
constant \cite{REFDN,REFANT}.
The action for this theory is akin to a Maxwell action except that the
$1$-form
gauge field is replaced by the $3$-form so that the field strength is a
$4$-form.  Whereas the action for this field depends on the metric on the
manifold, it does not describe any continuous degrees of freedom.
There is only the one degree of freedom over all space (not one per point)
given by the integration constant from the solution of the  equations of
motion.
This method of obtaining the cosmological constant differs from the usual
avenues
used for generating it \cite{REFW} as  no dynamics for
the field beyond this integration constant is possible.

The purpose of this paper is to pursue the  issue of the non-perturbative
quantum behaviour of such theories.
For the most part,  we will work with spacetimes of arbitrary dimensions,
$D$. We will thus study (D-1)-form ($\ca$) theories in $D$
dimensions\footnote{
Interestingly, in three dimensions,
$\ca$ is a 2-form or rank-two anti-symmetric tensor and it appears
in the low-energy effective field theory of three dimensional strings.}.
Due to the fact that a $\dimm$-form in $\dimm$ dimensions is dual to a scalar
field,
the kinetic lagrangian is a coordinate scalar which is constructed without a
metric.
Thus the action is invariant under volume-preserving diffeomorphisms.
As a result of this, we expect to find the metric appearing in terms of the
volume of spacetime in our quantum computations.
Our focus will be on three points:  The first is whether or not the vacuum
expectation value of the energy-momentum tensor is that of a cosmological
constant.  Secondly, these theories contain interesting geometrical
information.  Indeed, since the theories are volume-preserving diffeomorphism
invariant, we expect that there should be a class of observables which respect
this symmetry.  In fact, we will see that certain correlation functions are
equal
to intersection numbers on the spacetime manifold.  The third issue is the
parallel we will see between the quantum mechanics of these theories and the
trivial case of the point particle.
This theme will be made explicit in our cursory study of the thermodynamics of
the pure field theories for $\ca$.
Throughout this work, we
will take the spacetime manifolds, $M$, to be compact and orientable.

In the next sub-section, we introduce the actions for the classical
theory and discuss some pertinent issues regarding length dimensions and the
need for coupling constants.  This sub-section will also serve to introduce
our
notation.  We then start our study of the quantum theory
by examining the canonical quantization of these theories in section
\ref{CANON}.  Next, in section \ref{COVAR}, the covariant quantization is
carried out.  There, the  form of the partition functions in terms of
determinants will be
given via the method of resolvents \cite{REFSCH,REFSCHa}.  An explicit
three-dimensional example via
BFV quantization \cite{REFHT} will also be worked out.  Much of the remainder
of the
paper
will focus on
the computation of the vacuum value of the energy-momentum tensor,
$\ev{T_{ab}}$, and a class of observables.  To do this, we will give a
detailed
discussion of a  two-dimensional example in section \ref{TWODIM}.  There we
reinterpret
the Maxwell theory in two-dimensions as a classical theory of the cosmological
constant.  More importantly,
the results of past analyses of two-dimensional Yang-Mills/Maxwell theory
will be used to give an exact treatment of the quantum theory.  The
presence of non-trivial topological sectors, however, will lead to a
complicated dependence of $\ev{T_{ab}}$, on the volume of the Riemann surface.
In section \ref{OBSERVE}, the two-point function of a class of
operators will be shown to be equal to  the intersections numbers of one and
$D$-dimensional submanifolds.  The coupling of  $\ca$ to scalar fields will be
considered in section \ref{SCALAR}.
Finally, the general analytic nature of the thermodynamic relations obtained
from volume-preserving diffeomorphism invariant theories will be discussed in
section \ref{THERMO}.  We will also discuss the modifications of the usual
thermodynamics of scalar field theory as a result of the coupling to $\ca$.
A summary of the results is given in section \ref{SUMMARY}.

\vskip 0.5truein\setcounter{equation}{0}
\section{Pure Classical Theory}\label{CLASTHRY}
\forcepar

In this section, we will extend some of the classical physics of pure 3-form
theory
in four dimensions, as discussed in \cite{REFANT,REFDJ} to  $\dimm\geq2$
dimensions.  This action will be second order in derivatives.
This section will also serve the useful purpose of establishing our
conventions.  We will also introduce a first order action which yields the
same classical physics as the second order theory.
The former action will primarily be used from section \ref{OBSERVE} on.

\subsection{Second Order Action}
\forcepar

Let $\ca$ be a $\rank$-form on a compact and orientable $\dimm$-dimensional
manifold, $M$; that is $\ca\in\Lambda^{\dimm-1}(M)$. In addition, let $\cf$
denote
the exterior derivative of $\ca$,
\be
\cf\seqv d\ca\ \ .\label{EQNCF}
\ee
Let $M$ be equipped with a metric $g_{ab}$ through which the Hodge dual
`$~^*~$' is defined.
Since $\cf$ is a $\dimm$-form, it is dual to
a $0$-form and thus the wedge product $\cf\mywedge\dual\cf$ is  a $\dimm$-form
which
can be integrated over $M$.  As it is quadratic in the field, $\ca$, we use it
as the action which defines the classical $\rank$-form theory:
\be
S^{II}\seq \inv{\mu}\int_M \cf\mywedge\dual\cf\mi i{\Theta\over{2\pi}}
\int_M\cf \ \ .\label{EQNSII}
\ee
Added to this action will be
the Einstein-Hilbert action for the metric.  However, as the only purpose it
will serve is to
give the Einstein tensor for Einstein's equations in our discussion of the
cosmological constant, we will drop it throughout.  At this stage $\mu$ is a
real, positive coupling constant whose relevance we will soon discuss.   The
last term is analogous to $\theta$ terms in Yang-Mills theory.  Although its
role is purely
quantum mechanical, we have added it here for later use.  Other boundary terms
may be added to this action \cite{REFDJ}, however, we will not study their
effects
in this work.  Note that in two dimensions the last term in $S^{II}$ is the
first Chern class of a $U(1)$ bundle.

The action (\ref{EQNSII}) is invariant under the local transformation:
\be
\ca\ \to\  \ca \pls d\chi\ \ ,\label{EQNSYM}
\ee
where $\chi\in\Lambda^{\dimm-2}(M)$ is the parameter of the transformation.
This symmetry is analogous to the
Maxwell gauge transformation.  The equation of motion for the $\ca$ field is
\be
\delta \cf\seq0\qquad{\rm or}\qquad d\dual\cf\seq0\ \ ,\label{EQNEQM}
\ee
where $\delta$ is the adjoint of the exterior derivative.
Now since $\dual\cf$ is a $0$-form, the solution of this equation is
\be
\cf\seq 2\pi\kappa \omega\ \ ,\label{EQNSOL}
\ee
where $\omega$ is the volume form on $M$ and $\kappa $ is a  constant.
When the solution (\ref{EQNSOL}) is substituted into the energy-momentum
tensor arising from $S_{(0)}$:
\be T_{ab}\seq \frac{\delta S}{\delta g^{ab}}\seq -g_{ab}
\inv{2\mu}(\dual \cf)^2\ \ ,\label{EQNEM}\ee
it is found that the only
contribution to  Einstein's equations
is a cosmological constant determined by the square of $\kappa$.   This is the
only information carried by the $\cf$-field.
We can use Stokes' theorem to write
\be
\int_{\partial M} \ca \seq 2\pi \kappa  \Omega(M)\ \ ,\label{EQNSTOKES}
\ee
where $\Omega(M)$  represents the volume of $M$.
This equation says that a flux of $\ca$-field through the boundaries of
$M$ leads to a cosmological constant.  This is the only
information carried by the $\cf$-field and it appears as an integration
constant.

For proper interpretation as a contributor to the cosmological constant,
$\kappa $ should be the vacuum value of $\dual\cf$.  In the absence of the
$\Theta$ term,  the
action for $\ca$ is a perfect Gaussian in $\dual\cf$ and we are forced to
conclude
that $\ev{\dual\cf}=0$; thus $\kappa =0$.  To get around this, $\Theta$ must be
non-zero.  Note, however, that such a term in the action  does not change the
equations of motion.

Our quantum
expressions are expected to involve the volume of spacetime, $\Omega(M)$;
thus
it will be convenient to construct a dimensionless quantity out of
$\Omega(M)$.
We take the length dimensions of $\cf$ to be, $\ldim{\cf}=0$, while
$\ldim{\dual\cf}=-D$.    This means
we should assign to $\mu$, $\ldim{\mu}=-D$.  Based on this assignment, we
summarize the length dimensions of the various objects in our theory as:
\begin{center}
\begin{tabular}{||c|c||}\hline
{OBJECT}        &$\ell{\rm dim}$\\ \hline
$\omega$        &$\dimm$\\ \hline
$d$			&$0$\\ \hline
$\Omega(M)\equiv \int_M \omega$&$\dimm$\\ \hline
$\cf$           &$0$\\ \hline
$\dual\cf$           &$-\dimm$\\ \hline
$\mu$           &$-\dimm$\\ \hline
$\kappa$        &$-\dimm$\\ \hline
\end{tabular}
\end{center}
Apart
from
making $ \mu \Omega(M)$ dimensionless, this means that $\ldim{\int_M \cf}=0$.
Also,
$\ldim{\frac{\kappa^2}{\mu}}=-\dimm$ has dimensions of a cosmological
constant.

\subsection{First Order Action}
\forcepar

A first order form of the two-dimensional Maxwell action proved useful
\cite{REFWIT} in performing quantum calculations. A similar action may be
written in $\dimm$-dimensions.  Its form is
\be
S^I\seq  \int_M \big[ i\Phi \mywedge \cf \mi \inv{4}\mu  \Phi\mywedge\dual\Phi
\pls  \zeta\mywedge \psi \mi  i{\Theta\over 2\pi}  \cf\big]\ \
.\label{EQNSCF}
\ee
In writing this action a pair of  Grassmann odd fields,  $\zeta$ and
$\psi$ were introduced.  The form degree, Grassmann parity, etc. of the
various
fields are summarized in the following table.  \begin{center}
\begin{tabular}{||c|c|c||}\hline
{FIELD} &DEGREE&GRASSMANN PARITY\\ \hline
$\ca$   &$\rank$&even\\ \hline
$\Phi$  &$0$&even\\ \hline
$\zeta$ &$\rank$&odd\\ \hline
$\psi$  &$1$&odd\\ \hline
\end{tabular}
\end{center}
Our motivation for introducing this pair of Grassmann odd fields follows from
their usefulness when we later
deduce position independent observables.

As
a classical theory, $S^C$,  is equivalent to eqn. (\ref{EQNSII}). Indeed,
apart from the fact that $\zeta$ and $\psi$ vanish via their equations of
motion, the other two fields must satisfy
\be
\cf\seq - i\half \mu \dual\Phi\qquad{\rm and}\qquad d\Phi\seq
0\quad\Longrightarrow\quad \delta \cf\seq 0\ \ .\label{EQNEQMF}
\ee
Equation (\ref{EQNSOL}) is the solution for $\cf$.   $\Phi=
\frac{i4\pi}{\mu}\kappa$ is the solution to its equation of motion.
The energy-momentum tensor is
\be
T_{ab}\seq \frac{\mu}{8} g_{ab} \Phi^2\ \ ,\label{EQNEMF}
\ee
and is seen to reduce to (\ref{EQNEM}) after the use of (\ref{EQNEQMF}).

The invariance of this action under (\ref{EQNSYM}) is manifest.
The action is also invariant under the following rigid symmetry:
\be
[Q,\ca\rbrace\seq \zeta\ \ ,\qquad [Q,\psi\rbrace\seq i(-)^D d\Phi\ \
,\label{EQNQSYMF}
\ee
with $\Phi$ and $\zeta$ being invariant under $Q$ and $\psi$ is required to
vanish on $\partial M$.  Due to the presence of
two Grassmann-odd variables in the action, there is an additional symmetry:
\be
[\cs,\ca\rbrace\seq \dual \psi\ \ ,\qquad [\cs,\zeta\rbrace\seq i(-)^D \dual
d\Phi\ \
,\label{EQNQSYMFS}
\ee
when $\psi$ vanishes on $\partial M$.
Both $Q$ and  $\cs$ take the
$k$-forms into  $k$-forms.  Based on these transformations we can write down
the descent equations:
\bea
d(\half \Phi^2)&\seq& [Q,i(-)^{D+1}\Phi\mywedge\psi\rbrace\ \ ,\nen
d(\half \Phi^2) &\seq& [\cs,-i\Phi\mywedge\dual\zeta\rbrace\ \ ,\nen
d(i\Phi\mywedge\zeta) &\seq& [Q,\zeta\mywedge\psi\pls i\cf\mywedge\Phi\rbrace\
\
,\nen
d(i\Phi\mywedge\psi)&\seq& [\cs,\dual \zeta\mywedge\psi\mi  i(-)^D\Phi\mywedge
d\dual\ca\rbrace
\ ,\label{EQNDESFS}
\eea
Observe that the charges are  nilpotent and  hermitian.
Consequently, $\ev{\Phi^2(x)}$, is $x$-independent.
As before, if the action
(\ref{EQNSCF}) is to be a theory of the cosmological constant,
$\ev{\Phi^2(x)}$
must also be metric independent  so that the energy-momentum
tensor (\ref{EQNEMF}) is that of a cosmological constant.

\vskip 0.5truein\setcounter{equation}{0}
\section{Quantization}
Although the actions $S^I$ and $S^{II}$ are  classically equivalent, their
quantum mechanics differs primarily through the harmonic (zero-)
mode of $\Phi$, $\eta_0$.  On closed manifolds the latter
enters only in the $\Phi^2$ term.   Calling $\tilde\Phi$
the non-zero-mode part of $\Phi$ we find that it is this quantity which,  when
integrated
over,  yields a partition function which is equivalent to the second order
theory.  Although the Gaussian integral
over the zero-mode is meaningless at the level of the partition function, this
will not be the case when we look at the observables in the theory.
Furthermore, we will see that the canonical quantization of the two theories
differ due to $\eta_0$.  For manifolds with boundaries, $\eta_0$ also
contributes
to the
 $\int_M\Phi\mywedge\cf$ term. Consequently, the relationship between the
partition
functions is more involved.  We will examine this further during our discussion
of the
two dimensional theories in section \ref{TWODIM}.  In this section we set
$\mu=2$.  Let us  first sketch the
canonical
quantizations.

\subsection{Canonical Quantization}\label{CANON}
\forcepar
For the purposes of this section we will foliate $M$ as $M=\Sigma\times\IR$.
Call $t:\Sigma\times \IR\to \IR$ the time coordinate.
Let us first count the degrees of freedom in $\ca$.  We do this at a given
fixed time, say $t=0$.   Take $\lambda$ to be a vector field
such
that ${\cal L}_\lambda t=0$ where $\cal L$ is the Lie derivative on $M$.  At
fixed
time, define $\ca_0\equiv i_\lambda \ca\in \Lambda^{D-2}(\Sigma)$,
$\chi_0\equiv i_\lambda \chi\in\Lambda^{D-3}(\Sigma)$, ${\vec \ca}\equiv
\ca\in\Lambda^{D-1}(\Sigma)$
and ${\vec \chi}\equiv \chi\in\Lambda^{D-2}(\Sigma)$. With this notation,  the
symmetry (\ref{EQNSYM}) decomposes as\footnote{Our analysis is a
generalization
of that given
for 2-form theories in \cite{REFHT}.}
\bea
\ca_0 &\ \to\ &\ca_0\pls {\partial\over{\partial t}} \vec\chi'\ \ ,\nen
{\vec \ca}&\ \to\ &{\vec \ca}\pls d_\Sigma \vec \chi'\ \
,\label{EQNSYMRED}\eea
where $\vec \chi'= \vec \chi - \int_0 dt d_\Sigma\chi_0$
and  $d_\Sigma$ is the exterior derivative on $\Sigma$.
At fixed time, there are ${\rm dim}\Lambda^{D-1}(\Sigma)+{\rm
dim}\Lambda^{D-2}(\Sigma)$
components of $\ca$.
In the $\ca_0$ transformation, there are
${\rm dim}\Lambda^{D-2}(\Sigma)$
parameters.  While in the
$\vec \ca$ transformation, there are only $\sum_{k=2}^D (-1)^k{\rm
dim}\Lambda^{D-k}(\Sigma)$
parameters after accounting for exact terms.
Consequently, (D-1)-form theory in $D$ dimensions contains
\be
N_{dof}\seq \sum_{k=1}^D (-1)^{k+1} {\rm dim}\Lambda^{D-k}(\Sigma)\seq 0\ \
,\label{EQNDOF}
\ee
degrees of freedom.  This result is in concert with the classical solution
(\ref{EQNSOL}).
We will also see this  from the path integral  approach.

If we call the momenta conjugate to ${\vec \ca}$ and $\ca_0$, $\vec\Pi$ and
$\Pi_0$, respectively, we find the constraints
\bea
\Pi_0 &\ \approx\ & 0\ \ ,\nen
\cg\seq\delta_\Sigma \vec\Pi&\ \approx\ & 0\ \ .\label{EQNCONS}\eea
The last constraint is reducible.  The vanishing of the momentum conjugate to
$\ca_0$ implies that
this field may be gauged away.  This is accomplished by the use
of the symmetry in the first line of equation (\ref{EQNSYMRED})  which says
that
$\ca_0$ may be arbitrarily shifted at fixed times.
Setting $\ca_0=0$,
we find that the Hamiltonian density is given by the square of $\vec\Pi$.
The energy eigenstates  are of the form $\Psi_k(\vec\ca)=\exp{[ik\int_\Sigma
\vec \ca]}$.

Further analysis is best done with the dual form of the phase space
coordinate.
Let $\cv$ be the Hodge-dual of $\ca$ on $\Sigma$.  Then the symmetry
(\ref{EQNSYMRED}) allows us to set $d_\Sigma \cv=0$ or $\cv$ to be a constant
on
$\Sigma$.   This does not completely fix the gauge symmetry, however.  The
remaining symmetry, for which $\cv$ shifts by an arbitrary constant, may be
used
to bring the range of values of $\cv$ to a compact subspace.  In this way, the
eigenfunction $\Psi_k$ is normalizable for compact $\Sigma$.

The important distinction between the canonical quantization of the action
$S^{II}$
versus $S^{I}$ is in the treatment of the zero-modes upon which the
wavefunctions depend.  In the first order case, the $\Phi$ zero-mode cannot be
shifted as the $\Phi^2$ term in the action breaks the symmetry.  Thus the
eigenfunctions are defined over $\IR$ and cannot be normalizable, unlike the
second order case.
\subsection{Covariant Quantization}\label{COVAR}
\forcepar
We now turn to the path integral quantization of the theories.  As we saw in
the previous subsections, the symmetry (\ref{EQNSYM}) is
reducible.
Although, our first inclination might be to apply the BFV \cite{REFHT}
quantization procedure, advances in topological field theories have provided
us
with a more direct way of obtaining the determinantal form of the partition
function of theories with reducible symmetries.
The method of resolvents was invented by  Schwarz \cite{REFSCH} for precisely
this
purpose.
We will apply this method to our theories.  Later we will show,
in the specific cases of three  dimensions, how the BFV algorithm
matches Schwarz's method.
We will ignore the zero-modes  in the remainder of this section.  However,
we will analyze their effects in the specific computations in later sections.
As such, our covariant quantization will apply to both $S^I$ and $S^{II}$.

\subsubsection{Partition Function Via Resolvents}
\forcepar

Let us begin by recapping the key points in Schwarz's description of
resolvents.
Assume that we are given a quadratic action, $S[\Psi]=\ev{\Psi,\khat \Psi}$,
on the space, $\Gamma_0$, of fields, of which $\Psi$ is a member. Take
$\ev{,}$
to  be the inner product  and $\khat$ to be  a self-adjoint operator both
acting on $\Gamma_0$. Let $\Gamma_0$ be a subspace of a much larger space of

fields, $\Gamma$; label the generic subspace as $\Gamma_i$.  The existence of
symmetries such as (\ref{EQNSYM}), implies that there exists operators, $\ohat
i$, such that ${\ohat 1}:\Gamma_1\to \Gamma_0$ is a symmetry of $S$.  This is
seen as the statement that $\khat {\ohat 1}=0$.  The reducibility of our
action
(\ref{EQNSII}) translates into the existence of linear operators $\ohat i$
such that ${\ohat i}:\Gamma_i\to \Gamma_{i-1}$ and the operators are
nilpotent:
${\ohat{i-1}}{\ohat i}=0$.
The linear spaces $\Gamma_i$ along with the linear operators, $\ohat i$, form a
sequence called the resolvent.
\be 0 \sar \Gamma_N\oparr{\ohat N}\Gamma_{N-1}\oparr{\ohat
{N-1}}\Gamma_{N-2}\oparr{\ohat {N-2}}\cdots \ \Gamma_0\oparr{\khat}{\tilde
\Gamma}_0\sar 0\ \ ,\label{EQNSEQ}\ee
where ${\tilde \Gamma}_0$ is the space which is dual to $\Gamma_0$.
Taking the $\Gamma_i$ to be vector bundles over a compact manifold and using
the fact that the operators are nilpotent, we realize the resolvent as an
elliptic complex.

Our action, $S^{II}$,  is second order in derivatives and falls under the
heading of
elliptic resolvents of the {\it first} kind \cite{REFSCHa}.
Following Schwarz, the Laplacians,
\bea
\dal_0&\seq& \khat \pls {\ohat 1}{\ohat 1}^\dagger\ \ ,\nen
\dal_i&\seq& {\ohat i}^\dagger{\ohat i} \pls {\ohat {i+1}}{\ohat
{i+1}}^\dagger
\ ,\qquad\hbox{for } i=1,\ldots,N\ \ ,\label{EQNLAP}\eea
may be constructed.
Then, under the assumption that (\ref{EQNSEQ}) is exact, he found that the
partition function is formally
\be
Z_M\seq \prod_{i=0}^N {\det}^{\alpha_i}[\dal_i]\ \ ,\qquad \alpha_i\seq
(-1)^{i+1} (i+1)/2 \ \ ,\label{EQNZR}
\ee
neglecting zero-modes.
This is also expressed as
\be
Z_M\seq {\det}^{-1/2}[\khat]\prod_{i=1}^N {\det}^{(-1)^{i+1}/2}[{\ohat
i}^\dagger {\ohat i}]\ \ .\label{EQNZZ}\ee
Note that the pre-factor of this last expression gives the correct partition
function for the Gaussian field.
Elliptic resolvents of the second kind have been applied to Chern-Simons and
other such $BF$-theories;  see ref. \cite{REFBT}, for more detailed
descriptions
of this.
Of course, the determinants must be suitably regularized in the presence of
zero-modes.

Let us now apply this algorithm to determine the partition function of our
theory.
Apart from the issues mentioned at the end of the previous paragraph,
it is now straightforward to write down the answer.
Our resolvent is composed of almost the entire de Rham complex,
\be
0 \sar \Lambda^0(M)\oparr{d}\Lambda^1(M)\oparr{d}\cdots \
\Lambda^{D-1}(M)\oparr{\delta d}\Lambda^{D-1}(M)\sar 0\ \ .\label{EQNSEQDR}\ee
We find that $\dal_i$ is given by the laplacian on (D-1-$i$)-forms:
Using this in eqn.  (\ref{EQNZR}), we obtain
\be
Z_D(M)\seq \prod_{k=0}^D {\det}^{(-1)^k k/2}[\lap_{(k)}]\seq T_D(M)\ \
,\label{EQNTM}\ee
where $T_D(M)$ is the Ray-Singer torsion \cite{REFRS} of the de Rham complex
of the $D$-manifold, $M$.  As the latter is a manifold invariant, so is
$Z_D(M)$.
This is  in concert with the absence of degrees of freedom in the
$D$-dimensional theory.  In fact, if we count the number of degrees of
freedom
each determinant corresponds to, we find
\be
N_{dof}\seq \sum_{k=0} (-1)^{k+1} k {\rm dim}\Lambda^k(M)\seq
 \sum_{k=0}^D (-1)^{k+1} k {D\choose k}\seq 0\ \ .\label{EQNDOFZ}\ee

\subsubsection{BRST Quantization: A 3D Example}
\forcepar

We continue our study of the path integral quantization by working out the
BRST gauge fixed action for the  three dimensional theory.
As the BRST gauge fixing of reducible systems such as anti-symmetric tensor
theory in four dimensions are now textbook examples \cite{REFHT}, we simply
write
down the combined
gauge fixing and ghost action for the aforementioned symmetry:
\be
S_{GH+GF}=\int\big[ \alpha_0 \delta \ca\mywedge\dual{\delta \ca} +
{\bar \chi}\mywedge\dual{\lap\chi} +
{\bar \rho}\mywedge\dual{\lap\rho} + \sigma\mywedge\dual{\lap\sigma}\big]
\ \ .\label{EQNSGHGF}
\ee
The gauge choices $\dual{\delta \ca}=0$, $\dual{\delta \chi}=0$ and
$\dual{\delta
\bar\chi}=0$, have been made.  There are a number of gauge parameters which
enter into this expression.  We have judiciously chosen values for all of them
except for $\alpha_0$.  The fields which appear in this action fill out the
following pyramid.

\let\picnaturalsize=N
\def\picsize{1.75in}
\def\picfilename{vpdfig1.eps}
\ifx\nopictures Y\else{\ifx\epsfloaded Y\else\input epsf \fi
\let\epsfloaded=Y
\centerline{\ifx\picnaturalsize N\epsfxsize \picsize\fi
\epsfbox{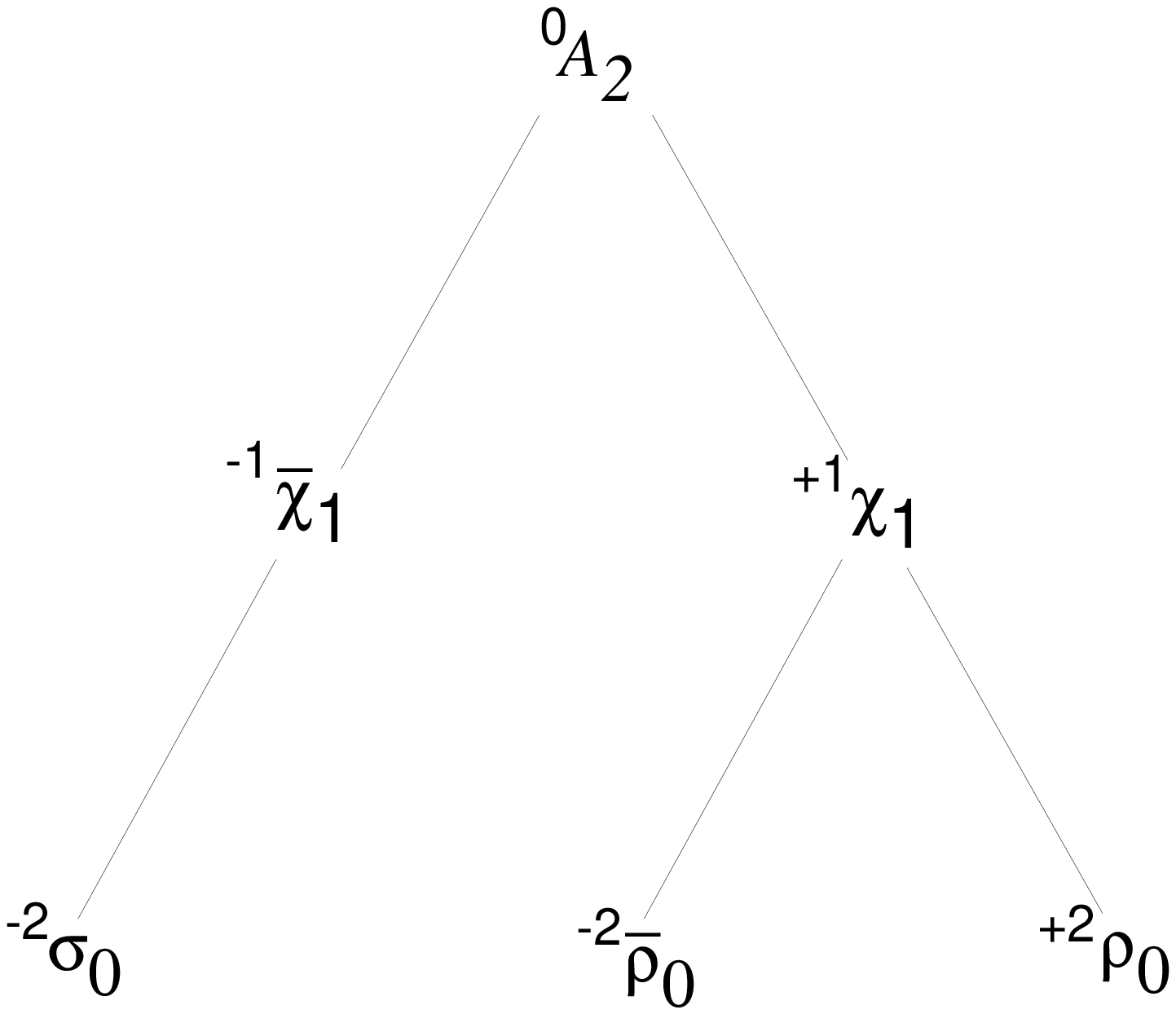}}}\fi
\centerline{\figfont Figure 1: The BFV pyramid for three dimensional $2$-form
theory.}

\noindent The notation $^{gh\#}\cc_k$ is used to indicate that the field $\cc$
is a $k$-form of the ghost number `${gh\#}$'.
The fields with odd ghost number are Grassmann odd.

When this action is added to (\ref{EQNSII}), the choice $\alpha_0=1$
leads to the total action
\be S_{\rm Tot}^{\rm local}=\int\big[ \ca\mywedge\dual{\lap\ca} +
{\bar \chi}\mywedge\dual{\lap\chi} +
{\bar \rho}\mywedge\dual{\lap\rho} + \sigma\mywedge\dual{\lap\sigma}\big]\ \
.\label{EQNTOTS}\ee
Thus, upon integrating over the fields we arrive at
\bea
Z_3(M)&\seq& {\det}^{-1/2}[\lap_{(2)}]
{\det}[\lap_{(1)}]{\det}^{-1}[\lap_{(0)}]{\det}^{-1/2}[\lap_{(0)}]\ \ ,\nen
&\seq& {\det}^{-3/2}[\lap_{(0)}]{\det}^{1/2}[\lap_{(1)}]\ \ ,\nen
&\seq&T_3(M)\ \ ,\label{EQNZ3}
\eea
neglecting the zero-modes.
We have thus found agreement with our previous derivation using the method of
resolvents.

\vskip 0.5truein\setcounter{equation}{0}
\section{Two Dimensions}\label{TWODIM}
\forcepar

In two dimensions, $\ca$ is a $1$-form, the Maxwell connection.
As is well known,  the local Maxwell symmetry given by shifts of $\ca$ by
locally exact
$1$-forms, leads to the absence of continuous degrees of freedom from this
two-dimensional action.  The results of subsection \ref{COVAR} may be viewed
as a generalization of this statement.  Numerous studies of two dimensional
Yang-Mills theory have appeared in the literature
\cite{REFMR,REFMAN,REFRAJ,REFF}.
Maxwell theory is seen to
be a simple example of these investigations.
Furthermore, the role of the Maxwell action as a contributor to the
Wheeler-DeWitt equation has been investigated in ref. \cite{REFBS}.
Thus, in this section, we will
extract various exact results from the literature and apply them to our
discussion.

As an example of the solution (\ref{EQNSOL}),
consider $S^2\simeq \IC P^1$ with Fubini-Study metric given by the Kahler form
$K=i\half \partial {\bar\partial} \ln(1+z{\bar z})=\frac{rdr\mywedge
d\theta}{(1+ r^2)^2}$ where $z=re^{i\theta}$, so that with $\cf=2\pi \kappa
K$,
$\int_{S^2} \cf= \pi$ or $\kappa=\half$.
We can give a more geometrical description of this classical solution.  The
integral $\int_M \cf^{(k)} = 2 \pi k$
defines the first Chern Class of the $U(1)$ bundle with curvature $\cf^{(k)}$.
The ``monopole'' charge $k$ is an integer.  $\cf$ is an element of
$H^2(M,\IR)$,
which is locally exact.

We  find it convenient to parametrize a solution of
Maxwell's equations as a superposition of the first Chern Classes which
satisfy
eqn. (\ref{EQNSOL}), $\cf=\sum_k c_k \cf^{(k)}$ then
$\kappa=\sum_k c_k
k$.
By following the analysis used in computing the partition function in
\cite{REFWIT,REFBT}, we can determine the values of the $c_k$ assumed by the
vacuum.  Then we will explicitly compute the vacuum value of the
energy-momentum tensor.

Let us
first recall the computations of the partition function of $S^{II}$.
To begin,
we set $\Theta=0$.  As the action is volume-preserving diffeomorphism
invariant
its partition function will depend only on the dimensionless quantity
$\epsilon\equiv \half \mu \Omega(M)$.   For an arbitrary Riemann surface $M$
it was found that
\bea
Z^{II}({\epsilon})& \seq &\sum_k Z^{II}_{k}({\epsilon})\nen
Z^{II}_{k}({\epsilon})& \seq &\pathint[d\ca] e^{-S^{II}} \delta (\int_M\cf
- 2\pi k)\ \ ,\label{EQNZ0}\eea
where the delta functional restricts the path integral to the gauge fields
with
first Chern Class $k$.  Observe that since $\inv{2\pi}\int_M \cf$ is a
manifold invariant, the delta function does not depend on the metric on
$M$.  Upon gauge fixing the $U(1)$ symmetry and introducing appropriate
Nicolai maps it is
possible to show that the various Jacobians cancel so that
\be  Z^{II}_{k}({\epsilon})\seq\sqrt{\frac{2\pi}{\epsilon}} e^{-2\pi^2
\frac{k^2}{\epsilon}}\ \ ,\label{EQNZq}\ee
and when written in terms of Jacobi theta functions (our conventions are those
of ref.  \cite{REFWW}),
\be  Z^{II}({\epsilon})\seq
\sqrt{-i\tau}\vartheta_3(0|\tau)\seq \vartheta_3(0|\tq)\ \
,\label{EQNZ}\ee
where
\be
\tau\equiv \frac{i2\pi}{\epsilon}=\frac{i4\pi}{\mu \Omega(M)}\ \ .\ee
The last equality in (\ref{EQNZq}) follows from the first after Poisson
re-summation.
This partition function has also been computed by a number of other methods;
see, for example refs. \cite{REFWIT,REFMR}.
  Observe that
$Z_{k}^{II}=Z_{-k}^{II}$ and $(-i\tau)$ is real and positive.

We can now compute correlation functions.  In order to extract the $c_k$, we
need to know $\ev{\int_M \cf}_{II}$.  It is simple to extend the
computation of this to
$\ev{(\int_M \cf)^n}_{II}$, for some positive integer $n$.  We find
\bea
\ev{(\int_M \cf)^n}_{II}& \seq & \inv{Z^{II}}\sum_k\pathint[d\xi]
\big(\int_M
\xi\big)^n
e^{-\inv{\mu}\int_M \xi\wedge \dual\xi}\delta (\int_M \xi -2\pi
k)\nen
& \seq & \inv{Z^{II}}\int d\nu \sum_k\pathint[d\xi]
e^{-\inv{\mu}\int_M\xi\wedge
\dual\xi}
(\inv{i}\frac{\partial}{\partial \nu} \pls 2\pi k)^n~\times\nen
&&\qquad\qquad\qquad\qquad\times ~e^{i\nu(\int_M \xi
\mi
2\pi k)}\nen
&\seq &\frac{(2\pi)^n}{Z^{II}}\sum_{k\in\bf Z} k^n Z^{II}_{k}({\epsilon})\ \
,\label{EQNVEV0}
\eea
which is zero for  $n$ odd; in particular, $\ev{\int_M\cf}_{II}=0$.
Note, that were we to compute this expectation value restricted to a
given Chern class with charge $k$, we would obtain $\ev{(\int_M
\cf)^n}_{II,k}=(2\pi
k)^n$.  It is then possible to give a representation of the $c_k$ coefficients
(see the discussion regarding eqn. (\ref{EQNSOL})) in the vacuum solution for
$\cf$
as the
the partition function for the charge $k$ normalized to the total $Z^{II}$:
\be  c_k\seq \frac{Z^{II}_{k}}{Z^{II}}\seq \frac{q^{k^2}}{\vartheta_3(0|\tau)}\
\
.\label{EQNck}\ee
The vacuum gets around the non-triviality of the complex line bundle by means
of this judicious choice of expansion coefficients, in forming its solution
for
$\cf$ via superposition.

A non-zero value for the vacuum solution
may be forced.
This is accomplished \cite{REFANT} with $\Theta\neq0$.
Via computations similar to those above but
involving
a shift of $\nu$ by $\Theta$,  it is possible to show that
\be  Z^{II}_{(\Theta)}({\epsilon})\seq \sum_k e^{i\Theta
k}Z_k^{II}({\epsilon})\seq
\sqrt{-i\tau} \vartheta_3(z|\tau)\ \
,\label{EQNZg}\ee
where $z\equiv\half\Theta$.
Notice that the difference between $Z^{II}$ and $Z_{(\Theta)}^{II}$ is a
partial phase
and that if $\inv{2\pi}\Theta\in {\bf Z}$ then
$Z_{(\Theta)}^{II}=Z^{II}$.   It is the non-integer values of
$\Theta\over{2\pi}$
which count. Then
$\ev{\int_M \cf}_{II}$ becomes
\be
\ev{\int_M \cf}_{II,\Theta}\seq\frac{2\pi}{Z_{(\Theta)}^{II}}\sum_{k\in\bf Z}
k
e^{i\Theta
k} Z_k({\epsilon})^{II}\seq -i\pi\frac{\partial \log
\vartheta_3(z|\tau)}{\partial z}\ \ .\label{EQNEVFF}\ee
This time the expansion coefficients are
\be
c_{(\Theta)k}\seq e^{i\Theta k} \frac{q^{k^2}}{\vartheta_3(z|\tau)}\ \
.\label{EQNck'}\ee
The distasteful aspect of this construction is that it defeats the original
purpose of obtaining the cosmological constant as a integration constant
independent of any constants in the action principle.  However, as we will see
below, it is best interpreted as a non-perturbative effect due to the
existence
of large gauge transformations.  In this context, its effect is to introduce
$k$ dependent relative phases
between the contributions from each first Chern Class.  In addition, it
has the form of a Wilson loop which in each topological sector is $e^{i\Theta
k}$.

We now compute the vacuum value of the energy-momentum tensor in the $S^{II}$

theory.  To do this, we use the definition of the energy-momentum tensor as a
functional derivative with respect to the metric to write
\bea
\ev{T_{ab}}_{II,\Theta}& \seq & \inv{Z_{\Theta}^{II}}\sum_k\pathint[d\ca]
T_{ab} e^{-S^{II}}\delta
(\int_M \cf \mi 2\pi k)\nen
&\seq &-\frac{\delta }{\delta  g^{ab}} \log Z_{\Theta}^{II}({\epsilon})\nen
&\seq &-\frac{\delta }{\delta  g^{ab}} \log \vartheta_3(z|\tq)\ \
.\label{EQNTEV}
\eea
It then follows that
\be \ev{T_{ab}}_{II,\Theta}\seq \inv{8\pi} \mu
g_{ab}\big[\half (-i\tau) \mi (-i\tau)^2 \frac{\partial}{\partial(-i\tau)}
\log{\vartheta_3(z|\tau)}\big]
\ \ .\label{EQNTVEV}\ee
Even with $\Theta\neq0$, there is a non-zero {\it vev} for $T_{ab}$.

Let us re-check our answer by direct calculation.  Returning to the explicit
expression for the energy-momentum tensor (\ref{EQNEM}),  we now compute the
vacuum value of $\ev{(\dual \cf)^2}_{II,\Theta}$.  This is given by
\bea
\ev{(\dual \cf(x))^2}_{II,\Theta}&\seq& \inv{Z_{\Theta}^{II}} \sum_k \pathint[d
\xi]
(\dual\xi(x))^2
e^{-\inv{\mu} \int_M \xi\mywedge \dual\xi\pls
i\frac{\Theta}{2\pi}\int_M\xi} \times\nen
&&\qquad\qquad\qquad\qquad\qquad\times~\delta(\int_M \xi -2\pi
k)\nen
&\seq& \inv{Z_{\Theta}^{II}} \big[\frac{\delta}{\delta(\dual J(x))^2} \int d\nu
\int
[d\xi] e^{-\inv{\mu} \int_M \xi\mywedge \dual\xi \pls
i\frac{\Theta}{2\pi}\int_M\xi}\times\nen
&&\qquad\qquad\qquad\qquad\qquad\times~e^{- \int_MJ\mywedge
\dual\xi} e^{i\nu(\int_M \xi -2\pi
k)}\big]{\mid}_{J=0}\nen
&\seq& -\frac{\mu}{2\Omega(M)} \pls
\frac{4\pi^2}{\Omega^2(M)}\frac{\sum_{k=1}^\infty k^2 \cos{[\Theta k]}
\exp{[-\frac{4\pi^2 k^2}{\mu \Omega(M)}]}}{\sum_{k=1}^\infty  \cos{[\Theta k]}
\exp{[-\frac{4\pi^2 k^2}{\mu \Omega(M)}]}}
\ \ .\nen
\label{EQNEVFSQ}
\eea
This is in agreement with equation (\ref{EQNTEV}) after use of  (\ref{EQNEM}).

The bracketed term in the
expression for $\ev{T_{ab}}_{II,\Theta}$ is not a constant, but depends on the
metric in a volume-preserving diffeomorphism invariant fashion.  The first term
in this expression (or the first term in eqn. (\ref{EQNEVFSQ})  is a zero-point
energy contribution.  In a theory with a fixed background it can be discarded.
With a dynamical metric, we then conclude  that $S^{II}$ is not a theory for
the cosmological constant.
The $\vartheta_3$ dependent  term arises from the non-triviality of
the $U(1)$ bundle over
$M$.  Notice that the result is genus independent.

\vskip 0.5truein\setcounter{equation}{0}
\section{Observables}\label{OBSERVE}
\forcepar
In this section we will compute the correlation functions of various
observables.  We first begin with the theory defined by the first order action,
$S^I$.  Here we will look at manifolds with and without boundaries.  Next, we
find one correlation function in the second order theory which contains only
global information.
\subsection{First Order Action}
\subsubsection{Manifolds Without Boundaries}
\forcepar
Thus far we have found the volume of spacetime to be ever present.  Since this
is
global information, we now ask if there are other global quantities which can
be computed in the quantum theory defined by $S^I$.  As the first part of this
action is of the form of a $BF$ theory, we wonder if any intersection numbers
survive as correlation functions in the $S^I$ theory.  In investigating this,
we will once again compute $\ev{T_{ab}}$.

Suppose we pick an integer, $N$, along with a set of integers $n_i$ and
locations
$x_i$ with $i=1,\ldots,N$.  Then we can define the correlation function
\be
\ev{\prod_{i=1}^N\Phi^{2n_i}(x_i)}_I\seq
\frac{\pathint [d\Phi][ d\ca]
\prod_{i=1}^N\Phi^{2n_i}(x_i)
e^{\int_M
\big[ i\Phi \mywedge \cf -\inv{4}\mu \Phi\mywedge\dual\Phi \big] }}
{\pathint [d\Phi][ d\ca ] e^{\int_M
\big[ i\Phi \mywedge \cf-\inv{4}\mu \Phi\mywedge\dual\Phi
\big] }}\ \ .\label{EQNPEV}
\ee
Since $\Phi$ only couples to itself and $\ca$, we need only concern
ourselves
with the integrals over these two fields as the others (ghosts, etc.)
will
cancel in the expectation value.

Given a complete set of eigenfunctions, $\phi_n(x)$, of the scalar laplacian,
$\triangle_{(0)}$, such that $\triangle_{(0)}\phi_n(x)=\omega_n^2\phi_n(x)$ and
$\int_M\phi_n\mywedge\dual\phi_m=\delta_{nm}$, for $n,m\neq0$, we expand
$\Phi$
as
\be
\Phi(x)\seq \big[\eta_0 \pls \sum_n{}' \eta_n \phi_n(x)\big]\ \
,\label{EQNPEXP}
\ee
where $\sum_n'$ means we have removed the zero-mode, $\eta_0$ from the sum.
Making use of eqn. (\ref{EQNQSYMF}) we find that
\be
\eta_n\seq (-)^D[Q,\inv{\omega_n^2} \int_M  \phi_n \mywedge
\dual
\delta\psi\rbrace\ \ ,\label{EQNXI}
\ee
for $n\neq0$.  The zero-mode, on the other hand,  cannot be written as
a $Q$-exact expression.
It then follows that the correlation function of $\Phi^k(x)$ with any operator,
$\co(y)$, which is $Q$ closed ($[Q,\co(y)\}=0$) is given in terms of  $\eta_0$:
$\ev{\Phi^k(x)\co(y)}_I=\ev{\eta_0{}^k\co(y)}_I$.  For a manifold
without
boundary or upon imposing the boundary condition that $\ca$ vanishes on
$\partial M$, it follows that eqn. (\ref{EQNPEV}) reduces to the simple
expression
\be
\ev{\prod_{i=1}^N\Phi^{2n_i}(x_i)}_I\seq \frac{\int_{-\infty}^{\infty} d\eta_0
\eta_0{}^{2p} e^{-\inv{2}
\epsilon\eta_0{}^2} }
{\int_{-\infty}^{\infty} d\eta_0  e^{-\inv{2}\epsilon\eta_0{}^2} }
\seq (2p-1)!! \epsilon^{-p}\ \
.\label{EQNPEVF}
\ee
Here, we used $\epsilon=\half \mu \Omega(M)$ as in section \ref{TWODIM} and
$p\equiv \sum_{i=1}^N n_i$.

Combing this result with equation (\ref{EQNEMF}) leads to
\be
\ev{T_{ab}}_I\seq  g_{ab}\inv{4 \Omega(M)}\ \ .\label{EQNEMFEV}
\ee
Observe that this final expression is independent of any coupling constants in
the action and the topology of $M$.

While obtaining this result, we dropped a surface term
$\eta_0\int_{\partial M} \ca$.  This is invalid if there are non-trivial
sectors
such as the first Chern class of the $U(1)$ bundle over a Riemann surface.
Consequently, we have only found the first term in the expansion of
(\ref{EQNTVEV}).  This is all we need, however, to establish the statement that
the
vacuum value of the energy-momentum tensors of the $(D-1)$-form field theories
are not of the form due to a  cosmological constant.

In going from the first to the second line in eqn.  (\ref{EQNTEV}) we assumed
that the measure was  metric independent.  We have also done so here.  Metric
dependence of the measure may arise from the regularization of the zero-modes
which in turn may lead to anomalies.  If such anomalies existed then they
should contribute to $\ev{T_{ab}}$.  In four dimensions, \cite{REFDN,REFCD},
the
integrated trace anomaly is found to be proportional to the Euler number on
$M$.  Summing this with the integral over $\ev{T_{a}{}^{a}}$, we do not find a
constant times $\Omega(M)$.  Thus our statement holds, anomalies
notwithstanding.

Thus far we have relied on the existence of the $Q$ or $\cs$ symmetry to
establish that
$d\ev{H(\Phi(x))}_I=0$ for some $C^1$ function $H(\Phi)$.  The use of these
symmetries was purely a convenience as we can demonstrate the same property
for
the action in which the Grassmann odd fields $\psi$ and $\zeta$ are absent
since
\bea
&\pathint& [d\ca][d\Phi] H'(\Phi(x)) d\Phi(x)e^{\int_M\big[ i\Phi\mywedge\cf
-\inv{4}\mu \Phi\mywedge\dual\Phi\big]}\nen
\seq &\pathint& [d\ca][d\Phi] H'(\Phi(x)) i \frac{\delta}{\delta
\ca(x)}e^{\int_M\big[ i\Phi\mywedge\cf-\inv{4}\mu
\Phi\mywedge\dual\Phi\big]}\nen
\seq &0&\ \ .\label{EQNAXIND}
\eea

The next set of volume-preserving diffeomorphism invariant observables we will
compute is given by the
generalized Wilson loops
\be
W(\theta)\seqv \exp{[i\theta\int_U\ca]}\ \ .\ee
Here $\theta$ is a parameter and $U$ is taken to be homologically trivial with
$U=\partial X$ where $X$
is a $D$-dimensional submanifold of $M$.  Write $W=\exp{[i\theta\int_X\cf]}$
and introduce a density
$J$ such that $\int_M J\mywedge f\equiv\int_X f$ for arbitrary
$D$-forms $f$; $W$ may then be expressed as
$W=\exp{[i\theta\int_MJ\mywedge \cf]}$.  For some integer, $N$,
of the Wilson
areas, $W_i$ ($i=1,\ldots,N$), we compute
\bea
\ev{\prod_{i=1}^N W_i(\theta_i)}_I\seq \exp\big[\mi&\frac{\mu}{4}&\sum_{i=1}^N
\frac{\theta_i^2 \Omega(X_i) \Omega(X_i')}{\Omega(M)}\nen
\mi &\frac{\mu}{2}&\sum_{i=1}^N\sum_{j<i}
\frac{\theta_i\theta_j\Omega(X_i)\Omega(X_j)}{\Omega(M)}\nen
\pls &\frac{\mu}{2}&\sum_{i=1}^N \sum_{j<i} \theta_i\theta_j \Omega(X_i\cap
X_j)~\big]\ \ ,
\eea
where $X'_i$ is the complement of $X_i$ in $M$.
This is found to be the straightforward generalization of the two-dimensional
results
of refs. \cite{REFWIT,REFBT}.  Once again, we see  the ubiquitous
dependence on
the
metric only through the volumes of submanifolds.  In this computation, only
trivial
bundles were assumed.

The final  correlation function we will compute is  the
two point function $\ev{\Phi(x)  \ca(y)}_I$.  The computation of $\ev{\Phi(x)
\ca(y)}$ follows from an extension of the
analysis
in refs. \cite{REFHS,REFBT}.  Use is made of the $\phi_n$ basis eigenfunctions
to define
$\ca_n \seq
-\omega_n{}^{-1}\dual d\phi_n$.  Upon expanding $\Phi$ and $\ca$ in terms of
these basis functions we find a simple form for the action,
\bea
\Phi(x)&\seq& \sum_n \eta_n \phi_n(x)\ \ ,\qquad \ca(x)\seq
\sum_n\alpha_n\ca_n\ \
,\nen
S^{I}&\seq& \sum_n S_n^I\ \ ,\qquad S_n^I\seqv\big[i\omega_n\eta_n\alpha_n\mi
\inv{4}\mu \eta_n{}^2\big]\ \